\PassOptionsToPackage{pdftex,dvipsnames}{xcolor}
\documentclass[submission, Phys]{SciPost}

\hypersetup{
	colorlinks,
	linkcolor={red!50!black},
	citecolor={blue!50!black},
	urlcolor={blue!80!black}
}

\usepackage[bitstream-charter]{mathdesign}
\usepackage{tikz,color}
\usetikzlibrary{arrows}

\DeclareSymbolFont{usualmathcal}{OMS}{cmsy}{m}{n}
\DeclareSymbolFontAlphabet{\mathcal}{usualmathcal}

\newcommand{\average}[1]{\left<{#1}\right>}

\begin{document}
\begin{center} \Large 
%\textbf{Synchronisation of two interacting particles in optical traps at different temperatures}
%\textbf{ Out-of-equilibrium dynamics of two hydrodynamically-coupled optically-trapped particles}

\textbf{ Out-of-equilibrium dynamics of two interacting optically-trapped particles}

%\alb{Concerning the title: why do we want to restrict to the  hydrodynamically-coupled  optically-trapped particles? Isn't our model more general?}
\end{center}
\begin{center}
 Victor S Dotsenko$^{1}$,  Alberto Imparato $^{2}$, Pascal Viot$^{1}$, Gleb Oshanin$^{1}$
\end{center}
\begin{center}
\em $^1$Sorbonne Universit\'e, CNRS, Laboratoire de Physique Th\'eorique de la Mati\`ere Condens\'ee (UMR CNRS 7600), 4 Place Jussieu, 75252 Paris Cedex 05, France\\
\em $^2$Department of Physics and Astronomy, University of Aarhus\\ Ny 
Munkegade, Building 1520, DK--8000 Aarhus C, Denmark
\end{center}

\begin{center}
	\today
\end{center}

\section*{Abstract}
We present a theoretical analysis of a non-equilibrium dynamics 
in a model system 
consisting of two  
particles 
which move randomly on a plane.  The two particles interact via a harmonic potential, 
experience their own (independent from each other) noises characterized by two different temperatures $T_1$ and $T_2$,  
and each particle is being held by its own optical tweezer. 
Such 
a system with two particle coupled by hydrodynamic interactions 
was previously realised experimentally in 
B\'erut et al. [EPL {\bf 107}, 60004 (2014)], and the difference between two temperatures has been achieved
by exerting an additional noise on either of the tweezers. Framing the dynamics
in terms 
 of two coupled over-damped Langevin equations, we show that the system reaches 
 a non-equilibrium steady-state with non-zero (for $T_1 \neq T_2$) probability currents that possess non-zero curls.
 As a consequence, in this system the particles are continuously spinning around their centers of mass in a completely synchronised way 
 - the curls of currents at the instantaneous positions of two particles have the same magnitude and sign.
 Moreover, we demonstrate that the components of currents of two particles are strongly correlated and undergo a rotational motion along closed elliptic orbits.

\vspace{2pt}

Keywords: Out-of-equilibrium dynamics, random spinning and rotational motion, molecular motors

\vspace{10pt}
\noindent\rule{\textwidth}{1pt}
\tableofcontents\thispagestyle{fancy}
\noindent\rule{\textwidth}{1pt}
\vspace{10pt}

\section{Introduction}
%When a system is in contact with two different thermal baths, it evolves towards
%a non equilibrium stationary state.  A minimal model for describing this situation consists in a system with  two degrees of freedom coupled by a bilinear interaction. First, Exatier and Peliti  \cite{Exartier1999} studied the response of two coupled spins with two temperatures. Fill and Reimann \cite{Filliger2007} considered a two-dimensional particle which exhibits a rotational motion due a a symmetry breaking. Many significant results of this modelalso called the Brownian gyrator,  have been obtained these last fifteen years in a series of paper\cite{Dotsenko2013,Ciliberto2013,Ciliberto2013a,Chiang2017,Fogedby2017,Chang2021,Mancois2018} and validated in several experiments
%\cite{Ciliberto2013a,Argun2017}.
%Solvable non equilbrium models  are useful to quantify the deviations to equilibrium by analysing the stochastic trajectories of particles\cite{Li2019}, and more generally for understandind the behavior of active matter\cite{Seifert2012,Marchetti2013,Gnesotto2018,Nascimento2021}.
%
%Different variants of the Brownian gyrator have been recently investigated\cite{Cerasoli2018,Sune2019a,Sune2019,Squarcini,Squarcini2022,Miangolarra2022,Squarcini2022b,Abdoli2021}. See also \cite{Dotsenko2019}
Within the recent years there was much interest in stochastic dynamics 
of out-of-equilibrium multicomponent systems,  different  components of which are connected to 
thermostats kept at different temperatures. On the theoretical side, several minimalistic (albeit experimentally-realisable) models have been worked out, providing deep insights into the general aspects of an emerging non-trivial 
and sometimes even a counterintuitive dynamical behavior. Such models  were also used as a  framework for checking the validity of various fluctuation relations and theorems \cite{Ruelle1999,Evans2002,Rondoni2007,Marconi2008,Sekimoto2010,Seifert2012,Ciliberto2017,Peliti2021}  and also for justifying the notion of effective temperatures \cite{Peliti2021,Puglisi2017}. A few stray examples are the Brownian gyrator model \cite{Exartier1999,Filliger2007}
 and its various generalizations \cite{Crisanti2012,Ciliberto2013a,Ciliberto2013,Dotsenko2013,Mancois2018,Fogedby2018,Bae2021,Cerasoli2018,Tyagi2020,Nascimento2021,Squarcini2022a,Cerasoli2022}, 
models of interacting particles connected to different heat baths \cite{Grosberg2015,Fogedby2017},
models of the directional
influence between cellular processes  \cite{Lahiri2017},
coupled Kuramoto oscillators kept at different temperatures \cite{Dotsenko2019}, bead-spring models \cite{Battle2016,Li2019,Imparato2021} and the molecular "spinning tops" in two-dimensional systems \cite{Dotsenko2022}. 
A common feature of several of these theoretical models is that they exhibit a motor effect, in the form of particle translational or rotational motion, as a consequence of both the broken spatial symmetry and the lack of thermal equilibrium.

On the experimental side, the behavior predicted by the theoretical analysis of the
Brownian gyrator model has been validated experimentally. It was done  by either constructing  equivalent 
electric circuits \cite{Ciliberto2013a,Ciliberto2013}, or by studying directly the dynamics of a Brownian 
colloidal particle that is optically trapped in an elliptical potential well and is simultaneously coupled to two
heat baths kept at different temperatures acting along perpendicular directions \cite{Argun2017}. Similarly, such out-of-equilibrium
systems were experimentally realised in a single-electron box consisting of two islands with a tunnel junction \cite{Koski2013} 
and  with two optically-trapped viscously coupled particles, in contact with two effective baths maintained at different temperatures \cite{Berut2014,Berut2016}.

The experimental set-up in \cite{Berut2014,Berut2016} consists of a disc-shaped cell (with $18 \, {\rm mm}$ in diameter and  
$1 \, {\rm mm}$ in depth) in which there are
two suspended micrometer-sized beads - $1$ and $2$ - that are confined by optical tweezers centered at two distinct 
spatial positions (see also \cite{Ziehl2009} for a similar set-up) at distance $15 \, {\mu m}$ above the lower surface of the cell and some distance $2 x_0$ apart of each other.  The two beads are experiencing two different effective temperatures - $T_1$ and $T_2$, respectively: this crucial condition is experimentally realised in  \cite{Berut2014,Berut2016}  by adding a Gaussian white noise to the position of either of the tweezers. As shown in \cite{Berut2014}, once the amplitude of the displacement is sufficiently small to ensure the validity of a linear regime, such an additional random force does not affect the stiffness of the tweezer (which therefore remains constant) but merely increases the effective temperature. Lastly, in such a set-up the beads are hydrodynamically coupled to each other; that being, they
interact between themselves through the motion of a surrounding viscous fluid. 
Formulating the model in terms of coupled Langevin equations for the positions of the beads and introducing the 
forces through the Rotne-Prager diffusion tensor, 
it was demonstrated in  \cite{Berut2014,Berut2016} (see also the earlier \cite{Ziehl2009} for the analysis  in the $T_1 = T_2$ case) that
the inter-bead interaction is elastic, i.e., is a quadratic function of the instantaneous distance between the beads, and the proportionally factor in this function is dependent in the leading order only on the fixed distance $2 x_0$ between the centers of the optical traps.  
A comparison of the solutions against an experimental data has shown that such an approximation is quite accurate. 
%\alb{I am not sure what you mean here, the forces in eq. 3 of ref.~\cite{Berut2016} are not linear in the particles' distance, which in that notation reads $x_1-x_2+d$.}
Clearly enough, this picture is only valid for sufficiently stiff traps such that the beads do not travel far away from the centers of their respective optical  traps. For "loose" traps this is not the case, and this is not the case either in situations when the distance $2 x_0$ 
becomes large and the beads get effectively decoupled from each other.

The theoretical analysis  in \cite{Berut2014,Berut2016} focused on the behavior of the effective heat fluxes between the two beads in the out-of-equilibrium state with $T_1 \neq T_2$.  It was demonstrated that these fluxes obey an exchange fluctuation theorem in the stationary state and moreover, the total hot-cold flux  satisfies a transient exchange fluctuation theorem at any time, while the total cold-hot flux obeys this theorem only at large enough times. However, these conceptually important results were derived under an assumption that the stochastic dynamics of the two particles can be viewed as an effectively \textit{one-dimensional} process that evolves along the line connecting the centers of two optical traps. Within such an assumption, the model becomes mathematically equivalent to the bead-spring model considered in \cite{Battle2016,Li2019} or the Brownian gyrator model with an external forcing \cite{Cerasoli2018,Cerasoli2021}.  Then, a legitimate question is whether 
due to such a restriction some remarkable features of the dynamical behavior are overlooked.  

In the present paper, motivated in part by the "spinning tops" model put forth in our recent paper \cite{Dotsenko2022}, we revisit the dynamical behavior in the system considered in \cite{Berut2014,Berut2016}, allowing now the beads to move on a  plane, which is somewhat closer to the actual geometrical set-up. Apart from the additional spatial dimension, our model here remains essentially the same as
the one formulated in \cite{Berut2014,Berut2016}: Each bead it optically trapped by its tweezer and the temperatures $T_1$ and $T_2$ at which the particles live are not equal to each other. We proceed to show that the dynamical behavior is indeed much more complex than in the 1D case :  In fact, it appears that the two beads undergo 
 a completely synchronized spinning around their centers of mass due to a systematic torque exerted on the particles. The term "completely synchronised" here means that not only the sign but also the magnitude of the curls of currents at the instantaneous positions $(x_1,y_1)$ and $(x_2,y_2)$ of the two particles  on a plane are exactly the same. Moreover, 
 examining the behavior of currents in a four-dimensional space $(x_1,x_2,y_1,y_2)$, we present an evidence that 
 the components of the currents of two particles are correlated and perform a rotational motion along closed elliptic orbits, which
 behavior resembles the dynamics of a Brownian gyrator \cite{Exartier1999,Filliger2007,Crisanti2012,Ciliberto2013a,Ciliberto2013,Dotsenko2013,Mancois2018,Fogedby2018,Bae2021,Cerasoli2018,Tyagi2020,Nascimento2021,Squarcini2022a,Cerasoli2022}. We stress that here, however, such a dynamical behavior is observed for the like components of currents of  the two particles (i.e., for the components $x_1 $ and $x_2$, or $y_1$ and $y_2$),  such that no net rotation of particles themselves around the origin or the centers of the traps takes place.

The paper is organized as follows: we introduce the model in Section \ref{sec:model}. Analytical expressions for the position probability density function and the probability currents in the steady-state are derived in section \ref{sec:stationary}. The synchronized spinning 
motion of the two particles is discussed in section \ref{sec:spinning}. The results of this section in the limit of a vanishingly small coupling parameter, in which limit they attain  a very compact form, are presented in Appendix \ref{ap}. 
Further on, the section \ref{sec:rotation} presents an analysis of the correlated behavior of currents in a four-dimensional space. We finally conclude in section \ref{sec:conclusion} with a brief recapitulation of our results.

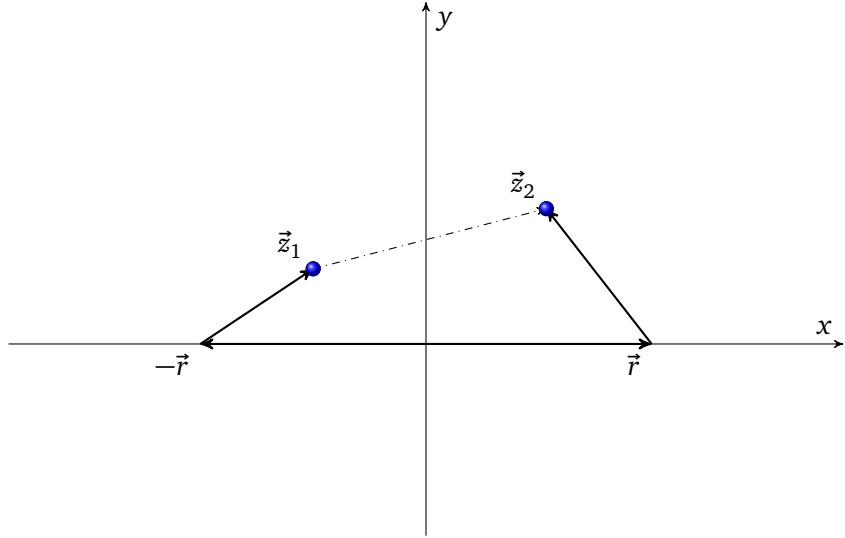
\begin{figure}[h]
	\begin{center}
		\begin{tikzpicture}
			[scale=1,auto, block3/.style   ={rectangle}]
			% Dimensions du repere
			\def\xmin{-5.55} \def\xmax{5.55} \def\ymin{-2.55} \def\ymax{4.55}
			% Styles des axes et de la grille
			\tikzstyle{axe}=[->,>=stealth']
			\tikzstyle{grille}= [step=1,very thin]
			% Grille
			%\draw [grille] (\xmin,\ymin) grid (\xmax,\ymax); 
			% Annotations axes et unites
			\draw [axe] (\xmin,0)--(\xmax,0) node[above left]  {$x$};
			\draw [axe] (0,\ymin)--(0,\ymax) node[below right] {$y$};
			\draw [axe,thick] (0,0)--(3,0) node[below left]  {$\vec{r}$};
			\draw [axe,thick] (0,0)--(-3,0) node[below left]  {$-\vec{r}$};
			\draw [axe,thick] (-3,0)--(-1.5,1) node[above left] {$\vec{z}_1$};
			\draw [axe,thick] (3,0)--(1.6,1.8) node[above left] {$\vec{z}_2$};
			%		\draw [axe,thick] (0,0)--(-1.5,1) node[above right] {$\vec{r}_1$};
			%		\draw [axe,thick] (0,0)--(1.6,1.8) node[above right] {$\vec{r}_2$};
			\draw [axe,dash dot] (-1.5,1)--(1.6,1.8) node[above right] {};%$\vec{r}_2$};
		\fill[ball color=blue] (-1.5,1) circle (0.1);;
		\fill[ball color=blue] (1.6,1.8) circle (0.1);;
		% Clip pour que les figures ne sortent pas du cadre
		\clip (\xmin,\ymin) rectangle (\xmax,\ymax); 
	\end{tikzpicture}
	\caption{A geometrical set-up of the two-dimensional model under study. Filled (blue) circles denote the particles $1$ and $2$. The vectors ${\bf r}$ and $-{\bf r}$ determine the centers of the fixed optical traps, while the vectors ${\bf z}_1$ and ${\bf z}_2$ show the instantaneous positions of the two particles on the $(x,y)$-plane, relative to the centers of their respective traps. The distance between the centers of the traps is fixed and equal to $2 x_0$.
	 }
	\label{fig:sketch}
\end{center}

\label{figure1}
\end{figure}

\section{The model}\label{sec:model}
Consider a two-dimensional system with two particles - $1$ and $2$, which are respectively confined by two optical tweezers centered at two distinct positions. Without a lack of generality, we assume that the centers of the traps are located  on the $x$-axis. We denote the positions of the centers of optical traps by vectors $-{\bf r}= (-x_0,0)$
and ${\bf r} = (x_0,0)$, which are both defined relative to the origin of the plane, and hence, the distance between the centers of the traps is fixed and equal to $2 x_0$.
In turn, the instantaneous positions of particles are 
specified by vectors ${\bf z}_1 = (x_1,y_1)$ and ${\bf z}_2 = (x_2,y_2)$, which are defined in the frames of reference centered at positions of the optical traps. According to such  a definition, these vectors therefore determine the displacements of respective particles from the centers of two potential wells. 

As shown in  \cite{Berut2014,Berut2016}, in realistic physical systems containing a solvent, 
the particles $1$ and $2$ are hydrodynamically coupled to each other - they
interact through the motion of a surrounding viscous fluid.  If the particles are sufficiently close to each other, the interaction potential $U(\rho)$  is a quadratic function of the inter-particle distance 
$\rho = |{\bf z}_1 - {\bf z}_2  + 2 {\bf r}|$,  
\begin{align} 
\label{u}
U(\rho) = \frac{u}{2} \rho^2 \,,
\end{align}
where $u$ is the constant coupling parameter (see \cite{Berut2014,Berut2016}). Lastly, due to the tweezers, the particles
are confined by the potential wells such that the overall potential energy $H({\bf z}_1, {\bf z}_2)$ is given by 
\begin{equation}\label{eq:ham}
	H\bigl({\bf z}_{1},{\bf z}_{2}\bigr) \; = \; 
	\frac{1}{2} \gamma \, {\bf z}_{1}^{2} \; + \; 
	\frac{1}{2} \gamma \, {\bf z}_{2} ^{2} \; + \; 
	\frac{1}{2} u \, \bigl({\bf z}_{1} - {\bf z}_{2}+2 {\bf r}\bigr)^{2}
\end{equation}
where the constant parameter $\gamma > 0$ defines the stiffness of the tweezers. 
We stress that in the physical situation considered in \cite{Berut2014,Berut2016} the form in Eq. \eqref{u} and hence, the total potential energy defined in Eq. \eqref{eq:ham}, 
are only valid for sufficiently small values of $x_0$.  For larger value of $x_0$ the hydrodynamic coupling between the particles vanishes and hence, Eq. \eqref{u} is no longer valid. Moreover, the parameter $\gamma$ should be sufficiently large such that the excursions of both particles away from the centers of their respective traps should be small, in order to ensure the validity of the form in Eq. \eqref{u}.
Having in mind these restrictions, we provide in what follows a formal solution of the model in Eq. \eqref{eq:ham} for arbitrary values of $\gamma > 0$ and arbitrary values of $u$, which may also attain negative values such that $u > -\gamma/2$. The meaning of the latter inequality will be made clear below.

%
%\begin{equation}
%	\label{2}
%	H\bigl({\bf z}_{1},{\bf rz}_{2}\bigr) \; = \; 
%	\frac{1}{2} (\gamma+u) \, {\bf z}_{1}^{2} \; + \; 
%	\frac{1}{2} (\gamma+u) \,{\bf z}_{2}^{2} \; - \; 
%	u \, {\bf z}_{1} {\bf z}_{2} \; - \; 2u  {\bf r} {\bf z}_{1} \; + \; 2u  {\bf r} {\bf z}_{2}
%	\; + \; 2u r^{2}
%\end{equation}
%%
We define next the dynamics of our model. 
Expanding the right-hand-side of Eq. \eqref{eq:ham} and dropping the constant term, which is irrelevant for the further analysis, we rewrite 
the total potential energy as 
\begin{equation}
	\label{3}
	H\bigl({\bf z}_{1},{\bf z}_{2}\bigr) \; = \; 
	\frac{1}{2} \kappa \, {\bf z}_{1}^{2} \; + \; 
	\frac{1}{2} \kappa \,{\bf z}_{2}^{2} \; - \; 
	u \, ({\bf z}_{1} \cdot {\bf z}_{2}) \; - \; 2u \, ({\bf r} \cdot {\bf z}_{1}) \; + \; 2u \, ({\bf r} \cdot {\bf z}_{2}),
\end{equation}
where the parameter $\kappa=\gamma + u$ and $(\,\, \cdot \,\,)$ denotes the scalar product.
Then, we stipulate that the deviations ${\bf z_1}$ and ${\bf z}_2$ of the particles positions from the centers of their respective traps obey a pair of coupled over-damped  Langevin equations:
\begin{eqnarray}
\nonumber
\frac{d}{dt} {\bf z_{1}}(t) &=& 
- {\boldsymbol \nabla}_{1} H\bigl({\bf z_{1}}, {\bf z_{2}}\bigr) 
\; + \; {\boldsymbol \xi}_{1}(t),
\\
\label{4}
\\
\nonumber
\frac{d}{dt} {\bf z_{2}}(t) &=& 
- {\boldsymbol \nabla}_{2} H\bigl({\bf z_{1}}, {\bf z_{2}}\bigr) 
\; + \; {\boldsymbol \xi}_{2}(t),
\end{eqnarray}
in which the symbols ${\boldsymbol \nabla}_{i}$ denote the gradient operators
while the vectors ${\boldsymbol \xi}_{i} \, = \, (\xi_{i}^{x}, \, \xi_{i}^{y}) \;, (i=1,2)$ 
stand for statistically-independent thermal noises,
with zero mean and the correlation function
\begin{equation}
\label{5}
\langle \xi^{\alpha}_{i}(t) \xi^{\beta}_{j}(t') \rangle \; = \; 
2 T_{i} \, \delta_{\alpha\beta} \, \delta_{ij} \; \delta(t-t'),
\end{equation}  
where $T_{1}$ and $T_{2}$ are the temperatures at which the particles $1$ and $2$ live.  
In the general case, $T_1 \neq T_2$, meaning that there is no unique temperature 
characterising the system and hence, the system does not converge to thermal equilibrium in the limit $t \to \infty$.  
We concentrate in what follows precisely on this out-of-equilibrium case 
seeking its consequences on the behavior of some observable properties. 

\section{Solution in the steady-state}\label{sec:stationary}

Let $P_t\bigr({\bf z_{1}}, {\bf z_{2}}\bigr)$ denote the position probability density function at time $t$ and
$P\bigr({\bf z_{1}}, {\bf z_{2}}\bigr)$ stand for its limiting form attained when $t \to \infty$. 
In this limit, the Fokker-Planck equation associated with the Langevin equations \eqref{4} has the form
\begin{equation}
\begin{split}
\label{6}
0 &= {\boldsymbol \nabla}_{1} 
\Bigl[
T_{1} {\boldsymbol \nabla}_{1} P\bigl({\bf z_{1}}, {\bf z_{2}}\bigr)
+ P\bigl({\bf z_{1}}, {\bf z_{2}}\bigr) 
{\boldsymbol \nabla}_{1} H\bigl({\bf z_{1}}, {\bf z_{2}}\bigr)
\Bigr]\\
&+ \; 
{\boldsymbol \nabla}_{2} 
\Bigl[
T_{2} {\boldsymbol \nabla}_{2} P\bigl({\bf z_{1}}, {\bf z_{2}}\bigr)
+ P\bigl({\bf z_{1}}, {\bf z_{2}}\bigr) 
{\boldsymbol \nabla}_{2} H\bigl({\bf z_{1}}, {\bf z_{2}}\bigr)
\Bigr]
\end{split}
\end{equation} 
Introducing the probability currents 
\begin{eqnarray}
\nonumber
{\bf j}_{1} \bigl({\bf z_{1}}, {\bf z_{2}}\bigr) &=&
T_{1} {\boldsymbol \nabla}_{1} P\bigr({\bf z_{1}}, {\bf z_{2}}\bigr)
+ P\bigl({\bf z_{1}}, {\bf z_{2}}\bigr) 
{\boldsymbol \nabla}_{1} H\bigl({\bf z_{1}}, {\bf z_{2}}\bigr),
\\
\label{7}
\\
\nonumber
{\bf j}_{2} \bigl({\bf z_{1}}, {\bf z_{2}}\bigr) &=&
T_{2} {\boldsymbol \nabla}_{1} P\bigr({\bf z_{1}}, {\bf z_{2}}\bigr)
+ P\bigl({\bf z_{1}}, {\bf z_{2}}\bigr) 
{\boldsymbol \nabla}_{2} H\bigl({\bf z_{1}}, {\bf z_{2}}\bigr),
\end{eqnarray}
one can conveniently rewrite 
the above Fokker-Planck equation \eqref{6} as
\begin{equation}
\label{8}
{\boldsymbol \nabla}_{1} \, {\bf j}_{1} \bigl({\bf z_{1}}, {\bf z_{2}}\bigr)
\; + \; 
{\boldsymbol \nabla}_{2} \, {\bf j}_{2} \bigl({\bf z_{1}}, {\bf z_{2}}\bigr)
\; = \; 0 \,, 
\end{equation}   
which implies that the total current is conserved.

Because the total potential energy in Eq. \eqref{3} 
is the quadratic function of the particles' positions, 
the solution is evidently a Gaussian function of the form
\begin{equation}
\label{9}
P\bigl({\bf z_{1}}, {\bf z_{2}}\bigr) \; = \; 
Z^{-1} \, 
\exp\Bigl(
-\frac{1}{2}\, \kappa \, A \, z_{1}^{2} \; - \; \frac{1}{2} \, \kappa \, B \, z_{2}^{2} 
\; + \; u C \, ({\bf z_{1}} \cdot {\bf z_{2}}) \; + \; 2u D \,  ({\bf r} \cdot {\bf z}_{1}) \; - \; 2u E ({\bf r} \cdot {\bf z}_{2})
\Bigr) \,,
\end{equation}
where $Z$ is a normalization constant, 
\begin{align}
\label{10}
Z \; &= \; \int\int d {\bf z}_{1} \, d {\bf z}_{2} \;  
\exp\Bigl(
-\frac{1}{2}\, \kappa \, A \, z_{1}^{2} \; - \; \frac{1}{2} \, \kappa \, B \, z_{2}^{2} 
\; + \; u C \, ({\bf z_{1}} \cdot {\bf z_{2}}) \; + \; 2u D  ({\bf r} \cdot {\bf z}_{1}) \; - \; 2u E ({\bf r} \cdot {\bf z}_{2})
\Bigr) \nonumber\\
&=\frac{4\pi^2}{AB\kappa^2-C^2u^2} \exp\left({\frac{2x_0^2u^2
		(\kappa(A E^2 +BD^2)-2u CDE)} {AB\kappa^2-C^2u^2}}\right) 
\end{align}
and the coefficients $A$, $B$, $C$, $D$ and $E$ are to be defined. In order to determine the unknown coefficients, 
we first substitute Eqs. \eqref{9} and \eqref{3} into Eqs. \eqref{7}, to get
the following expressions for the currents
\begin{eqnarray}
\label{11}
{\bf j}_{1} \bigl({\bf z_{1}}, {\bf z_{2}}\bigr)  &=&
\Bigl[
\kappa \,\bigl(1 - A T_{1} \bigr) \, {\bf z_{1}} \; + \; 
u \, \bigl( C T_{1}  - 1 \bigr) \, {\bf z_{2}} \; + \; 
2u \; \bigl(D T_{1}  - 1 \bigr) \, {\bf r}
\Bigr]
\;  P\bigl({\bf z_{1}}, {\bf z_{2}}\bigr)
\\
\nonumber
\\
\label{12}
{\bf j}_{2} \bigl({\bf z_{1}}, {\bf z_{2}}\bigr)  &=&
\Bigl[
\kappa \,\bigl(1 - B T_{2} \bigr) \, {\bf z_{2}} \; + \; 
u \, \bigl( C T_{2}  - 1 \bigr) \, {\bf z_{1}} \; - \; 
2u \; \bigl(E T_{2}  - 1 \bigr) \, {\bf r} 
\Bigr]
\;  P\bigl({\bf z_{1}}, {\bf z_{2}}\bigr) \,.
\end{eqnarray}
Inserting next the above expressions into the Fokker-Planck equation \eqref{8}, 
we obtain {\it six} equations for {\it five} unknown coefficients
$A$, $B$, $C$, $D$ and $E$: 
\begin{align}
 \label{14}
&A(1 - AT_{1}) - \frac{u^{2}}{\kappa^2}  \, C (CT_{2} \, - \, 1) \; = \; 0,
\\
&
B(1 - BT_{2}) - \frac{u^{2}}{\kappa^2}  \, C (CT_{1} \, - \, 1) \; = \; 0,
\label{15}
\\
&A(1-2 C T_1)+B(1-2 CT_2)=-2C,
\label{16}\
\\
&
 \frac{u}{\kappa}(E +C (1-2 E T_2))+(A+D-2 A D T_1)=0,
\label{17}
\\
&
\frac{u}{\kappa}(D +C (1-2 D T_1))+(B+E-2 B E T_2)=0,
\label{18}
\\
\label{19}&
\kappa (1-AT_{1}) \, + \, \kappa (1-BT_{2}) \, + \, 
2\frac{u^2x_0^2}{\kappa} (D(DT_{1}-1) \, + \,  E(ET_{2}-1)) \; = \; 0 .
\end{align}
From Eqs. \eqref{14} to \eqref{16}, we readily find that
$A$, $B$, and $C$ obey

\begin{align}\label{eq:A}
A&=\frac{1}{T_1}+\frac{ u^2 (T_1^2-T_2^2)}{(4 \kappa ^2 T_1 T_2+u^2 (T_1-T_2)^2)T_1},\\\label{eq:B}
B&= \frac{1}{T_2}-\frac{ u^2 (T_1^2-T_2^2)}{(4 \kappa ^2 T_1 T_2+u^2 (T_1-T_2)^2)T_2},\\\label{eq:C}
C&=\frac{2 \kappa ^2 (T_1+T_2)}{4 \kappa ^2 T_1 T_2+u^2 (T_1-T_2)^2}.
\end{align} 
Then,  Eqs. \eqref{17} and \eqref{18} give
\begin{align}\label{eq:D}
D=&\frac{4 \kappa^2T_2+2\kappa u(T_1-T_2) }{4 \kappa ^2 T_1 T_2+u^2 (T_1-T_2)^2},\\\label{eq:E}
E=& \frac{4 \kappa^2T_1+2\kappa u(T_2-T_1) }{4 \kappa ^2 T_1 T_2+u^2 (T_1-T_2)^2} \,.
\end{align} 
Note that for the above solution the Eq. \eqref{19} holds as an identity, so that the system of equations \eqref{14} to \eqref{19}  is not overdetermined, and also that the coefficients are actually independent of the distance $2 x_0$ between the centers of the optical traps, which enters only in Eq. \eqref{19}.

 We are now equipped with all necessary ingredients to find explicit expressions for the normalisation $Z$ and the probability currents.  Inserting Eqs. \eqref{eq:A} to \eqref{eq:E} into Eq. \eqref{10}, we have

\begin{equation}\label{eq:Z}
	Z=\frac{\pi^2((T_2-T_1)^2u^2+4T_1T_2\kappa^2)}{\kappa^4-\kappa^2u^2}
	\exp\left(\frac{8x_0^2(T_1+T_2)u^2\kappa^2}{(u+\kappa)((T_2-T_1)^2u^2+4T_1T_2\kappa^2)}\right)	
\end{equation}
Note that the normalization constant $Z$, Eq.\eqref{eq:Z}, is bounded when $\gamma > 0$ and positive whenever $\kappa^2>u^2$, in which case the system is stable. The latter inequality is realised when 
 $u > -\gamma/2$, (recall that $\kappa = u + \gamma$), which explains the above imposed constraint (see the paragraph below Eq. \eqref{eq:ham}). Note, as well, that  the parameter $u$ can therefore be negative meaning that our analysis 
 is also valid for the systems in which the particles (sufficiently weakly) repel each other. In turn, the probability currents are given explicitly by

\begin{equation} \begin{split} \label{eq:j1}
	&{\bf j}_{1} \bigl({\bf z_{1}}, {\bf z_{2}}\bigr)  =
  \frac{u (T_2-T_1)  P\bigl({\bf z_{1}}, {\bf z_{2}}\bigr)}{4 \kappa ^2 T_1 T_2+u^2 (T_1-T_2)^2} \\
&\;\; \times	\Bigl[
\kappa	u(T_1+T_2)  {\bf z_{1}} \; - \; 
	 ((T_2-T_1)u^2+2\kappa^2T_1) {\bf z_{2}}  \; + \; 
	2((T_2-T_1)u^2+2u\kappa T_1) \, {\bf r}
	\Bigr]
	\\	
	&{\bf j}_{2} \bigl({\bf z_{1}}, {\bf z_{2}}\bigr)  =
 \frac{u (T_1-T_2)  P\bigl({\bf z_{1}}, {\bf z_{2}}\bigr)}{4 \kappa ^2 T_1 T_2+u^2 (T_1-T_2)^2} \\
&\;\; \times		\Bigl[
\kappa	u(T_1+T_2)  {\bf z_{2}} \; - \; 
((T_1-T_2)u^2+2\kappa^2T_2) {\bf z_{1}}  \; + \; 
2((T_1-T_2)u^2+2u\kappa T_2) \, {\bf r}
\Bigr]
\end{split}
\end{equation}

Therefore, in out-of-equilibrium conditions (i.e. for $T_1 \neq T_2$), and also for a non-zero coupling between the two particles (i.e. when $u \neq 0$),
there exist non-vanishing probability currents in the steady-state. Below we discuss some remarkable features of the dynamical behavior, which originate from this latter circumstance.

\section{Synchronous spinning of particles}\label{sec:spinning}

Our aim now is to demonstrate that the probability currents possess a non-zero curl, i.e., the velocity field undergoes a circulation.
The curls $S_1({\bf z_{1}},{\bf z_{2}})$ and $S_2({\bf z_{1}},{\bf z_{2}})$  are formally defined as the circulation density at  "point" $({\bf z}_1,{\bf z}_2)$ of the field, i.e., $S_1({\bf z_{1}},{\bf z_{2}}) = ({\boldsymbol \nabla}_{1}\times {\bf j}_{1})\cdot {\bf \hat k}$ and $S_2({\bf z_{1}},{\bf z_{2}}) = ({\boldsymbol \nabla}_{2}\times {\bf j}_{2})\cdot {\bf \hat k}$, where
${\bf \hat k}$ is the unit vector in the direction orthogonal to the $(x,y)$-plane and the symbol $(\,\, \times \,\,)$ denotes the vector product.  Taking advantage of the above equations \eqref{eq:j1}, we readily find that the curls are given explicitly by
%\begin{eqnarray}
%S_{1}\bigl({\bf z_{1}},{\bf z_{2}}\bigr) &=&({\boldsymbol \nabla}_{1}\times {\bf j}_{1})\cdot {\bf \hat k}=\frac{2 (T_1 - T_2) u (u - \gamma) \gamma (2 r u (y_1 + y_2) + (x_2 y_1 - x_1 y_2) (u + \gamma)}{(T_1 - T_2)^2 u^2 + 4 T_1 T_2 \gamma^2}P\bigl({\bf z_{1}}, {\bf z_{2}}\bigr)\\
%S_{2}\bigl({\bf z_{1}},{\bf z_{2}}\bigr) &=&({\boldsymbol \nabla}_{2}\times {\bf j}_{2})\cdot {\bf \hat k}=\frac{2 (T_1 - T_2) u (u - \gamma) \gamma (2 r u (y_1 + y_2) + (x_2 y_1 - x_1 y_2) (u + \gamma)}{(T_1 - T_2)^2 u^2 + 4 T_1 T_2 \gamma^2}P\bigl({\bf z_{1}}, {\bf z_{2}}\bigr),
%\end{eqnarray} 
\begin{align}
\label{S1}
S_{1}\bigl({\bf z_{1}},{\bf z_{2}}\bigr)  =\frac{2 u \lambda \kappa (T_1 - T_2)  \left[2 x_0 u (y_1 + y_2) + (x_2 y_1 - x_1 y_2) (u + \kappa)\right]}{(T_1 - T_2)^2 u^2 + 4 T_1 T_2 \kappa^2}P\bigl({\bf z_{1}}, {\bf z_{2}}\bigr) \,,
%S_{2}\bigl({\bf z_{1}},{\bf z_{2}}\bigr) &=&({\boldsymbol \nabla}_{2}\times {\bf j}_{2})\cdot {\bf \hat k}=\frac{2 (T_2 - T_1) u (u - \kappa) \kappa [2 r u (y_1 + y_2) + (x_2 y_1 - x_1 y_2) (u + \kappa)]}{(T_1 - T_2)^2 u^2 + 4 T_1 T_2 %\kappa^2}P\bigl({\bf z_{1}}, {\bf z_{2}}\bigr),
%S_{2}\bigl({\bf z_{1}},{\bf z_{2}}\bigr) &=S_{1}\bigl({\bf z_{1}},{\bf z_{2}}\bigr)
\end{align} 
and
\begin{align}
\label{S2}
S_{2}\bigl({\bf z_{1}},{\bf z_{2}}\bigr)  =\frac{2 u \lambda \kappa (T_1 - T_2)  \left[2 x_0 u (y_1 + y_2) + (x_2 y_1 - x_1 y_2) (u + \kappa)\right]}{(T_1 - T_2)^2 u^2 + 4 T_1 T_2 \kappa^2}P\bigl({\bf z_{1}}, {\bf z_{2}}\bigr) \,.
\end{align}
Remarkably, the curls $S_1({\bf z_{1}},{\bf z_{2}})$ and $S_2({\bf z_{1}},{\bf z_{2}})$ are a) both non-zero in out-of-equilibrium conditions and for $u \neq 0$ and moreover, b) are exactly equal to each other at any point $({\bf z_{1}},{\bf z_{2}})$.
First, this implies that if the particles were to have a finite-size, the field will create
 a net torque on each particle such that it will steadily spin about its center of mass. Second, such a spinning motion of the two particles will be completely synchronized in the sense that both the sign and the magnitude of the curls $S_1({\bf z_{1}},{\bf z_{2}})$ and $S_2({\bf z_{1}},{\bf z_{2}})$ are exactly the same. For $u > 0$ and $T_1 > T_2$, the curls will be positive if the coordinates of particles' displacements from the centers of the optical traps obey
 \begin{align}
 \label{m}
 2 x_0 u (y_1 + y_2) + (x_2 y_1 - x_1 y_2) (u + \kappa) > 0 \,,
 \end{align}
 and will be less than zero, otherwise.  When Eq. \eqref{m} becomes an equality, the curls vanish such that the spinning motion stops. This happens, in particular, when both particles appear at the centers of their respective optical traps.

The curl of either of the currents, e.g., of ${\bf j}_{1} \bigl({\bf z_{1}}, {\bf z_{2}}\bigr)$, integrated over all possible positions of either of the particles vanishes, i. e., 
\begin{align}
\int d{\bf z}_1 \, S_{1}\bigl({\bf z_{1}},{\bf z_{2}}\bigr) = \int d{\bf z}_2 \, S_{1}\bigl({\bf z_{1}},{\bf z_{2}}\bigr) = 0 \,.
\end{align}
By symmetry, the same is true for $S_{2}\bigl({\bf z_{1}},{\bf z_{2}}\bigr)$. It seems interesting, however, to determine a property which does not vanish when it is integrated over positions of the particles. To this end, we consider 
  the absolute values 
  of the curls integrated over all possible positions of one of the particles 
  with the second one being fixed  at the center of the optical trap :
\begin{equation}
\begin{split}
\label{def}
	\average{|S_1|}=& \int d{\bf z_{2}} |S_1(0, {\bf z_{2}})|\\
	\average{|S_2|}=& \int d{\bf z_{1}} |S_1( {\bf z_{1}},0)| \,.
	\end{split}
 \end{equation} 
Inserting our expressions \eqref{S1} and \eqref{S2} into Eqs. \eqref{def} and performing the integrations, we find after some algebra
\begin{equation}
\label{nu}
\begin{split}
\average{|S_1|}
%&=\frac{4 \kappa  r u^2 |(T_2-T_1)| (u-\kappa )^2 (\kappa +u)  }{\pi ^{3/2} \left|u^2 (T_1-T_2)-2 \kappa ^2 T_1\right|%^2}\sqrt{\frac{\kappa  u^2 (T_2-T_1)+2 \kappa ^3 T_1}{4 \kappa ^2 T_1 T_2+u^2 (T_1-T_2)^2}} \E^{\frac{4 \kappa  r^2 u^2 %(u-\kappa )}{(\kappa +u) \left(u^2 (T_2-T_1)+2 \kappa ^2 T_1\right)}} \nonumber \\
&=\frac{4 \kappa (\gamma +2u) x_0 u^2 \gamma^2  |(T_2-T_1)|    }{\pi ^{3/2}  \sigma_1^3 }\sqrt{\frac{(u + \gamma) }{  (T_1 + T_2)^2 u^2 + 8 T_1 T_2 u \gamma + 4 T_1 T_2 \gamma^2} } \\
&\times \exp\left(\frac{-4 (u+\gamma) \gamma x_0^2 u^2}{(\gamma +2 u) \sigma_1^2}\right),\\
\average{|S_2|}
%&=\frac{4 \kappa  r u^2 |(T_2-T_1)| (u-\kappa )^2 (\kappa +u)  }{\pi ^{3/2} \left|u^2 (T_1-T_2)+2 \kappa ^2 T_2\right|%^2}\sqrt{\frac{\kappa  u^2 (T_1-T_2)+2 \kappa ^3 T_2}{4 \kappa ^2 T_1 T_2+u^2 (T_1-T_2)^2}}\E^{\frac{4 \kappa  r^2 u^2 %(u-\kappa )}{(\kappa +u) \left(u^2 (T_1-T_2)+2 \kappa ^2 T_2\right)}}\nonumber \\
&=\frac{4 \kappa (\gamma +2u) x_0 u^2 \gamma^2  |(T_2-T_1)|    }{\pi ^{3/2}  \sigma_2^3 }\sqrt{\frac{(u + \gamma) }{  (T_1 + T_2)^2 u^2 + 8 T_1 T_2 u \gamma + 4 T_1 T_2 \gamma^2} }\\
&\times \exp\left(\frac{-4 (u+\gamma) \gamma x_0^2 u^2}{(\gamma +2 u) \sigma_2^2}\right),
\end{split} 
\end{equation}
where we have used the shortened notations
\begin{align}
\sigma_1^2&\equiv(T_1 + T_2) u^2 + 4 T_1 u \gamma +  2 T_1 \gamma^2,\\
\sigma_2^2&\equiv(T_1 + T_2) u^2 + 4 T_2 u \gamma +  2 T_2 \gamma^2\,.
\end{align} 
Hence, the integrated absolute values  of the curls $\average{|S_1|}$ and $\average{|S_2}|$ do not vanish when the product  $x_0 u(T_2-T_1)\neq 0$. This occurs when the following three conditions are  simultaneously met: the temperatures are different, the coupling between particles and also the distance between the two optical centers is not equal to zero.  The non zero values of $\average{|S_1|}$ and $\average{|S_2|}$ imply that there exists a synchronized motion of particles in the stationary state.

It may be also instructive to consider the ratio of  $\average{|S_1|}$ and $\average{|S_2}|$. From Eqs. \eqref{nu}
 we find 
\begin{equation}
\frac{\average{S_1}}{\average{S_2}}= 
\exp\left(\frac{8 \kappa  x_0^2 \lambda^2 u^2 (T_1-T_2)}{4 \kappa ^4 T_1 T_2-u^4 (T_1-T_2)^2+ \kappa ^2 u^2 (T_1-T_2)^2}\right)   
\left|\frac{u^2 (T_2-T_1)+2 \kappa ^2 T_2 }{u^2 (T_1-T_2)-2 \kappa ^2 T_1}\right|^{3/2}.
\end{equation} 
Expanding the latter expression in powers of the coupling parameter $u$, we have
\begin{equation}
\label{zz}
\frac{\average{|S_1|}}{\average{|S_2|}}=\left(\frac{T_2}{T_1}\right)^{3/2}+\frac{u^2}{4 \kappa ^2 } \frac{(T_1 - T_2)}{T_1 T_2}  \left(\frac{T_2}{T_1}\right)^{3/2} \left(8 \kappa  x_0^2+3 T_1+3 T_2\right) +O\left(u^3\right) \,,
%-\frac{4 u^3}{\kappa ^2} r^2 (1/T_2-1/T_1) \left(\frac{T_2}{T_1}\right)^{3/2}+O\left(u^4\right)
\end{equation} 
where the symbol $O\left(u^3\right)$ signifies that the omitted correction terms are proportional to  $u^3$. 
Equation \eqref{zz} implies
 that $\average{|S_1|}$ and $\average{|S_2}|$ can be disproportionally different, if the temperatures are very different. In particular, 
$\average{|S_1|}$ can be much larger than $\average{|S_2}|$ if $T_2 \ll T_1$.
%Therefore, in the limit of a small coupling between the two particles, the intensity of the synchronization 
%is larger for the cold particle than for the hot particle.

\section{Correlated behavior of currents}\label{sec:rotation}

In this section we discuss an emerging cooperative behavior of the probability currents 
defined in Eqs. \eqref{eq:j1}. To this end, we study the correlations between the components of the probability currents in a four-dimensional space 
$(x_1,x_2,y_1,y_2)$, accessing them via the streamplots of the projections on different planes. 

\begin{figure}[h]
	\center
	\includegraphics[width=6cm]{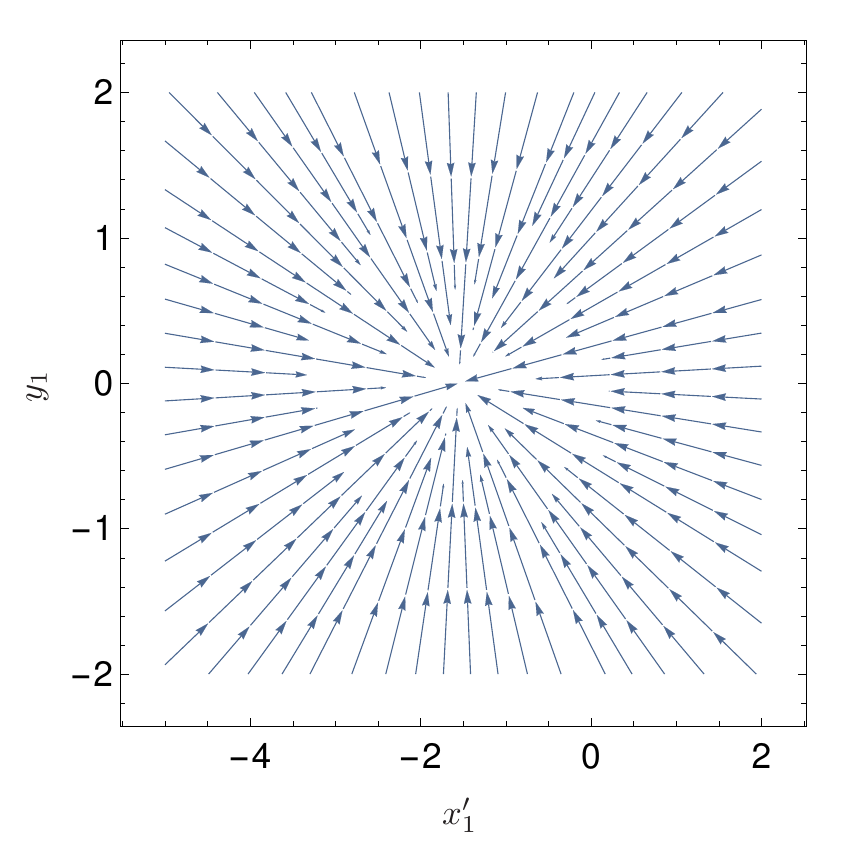}
	\hspace{.75cm}\includegraphics[width=6cm]{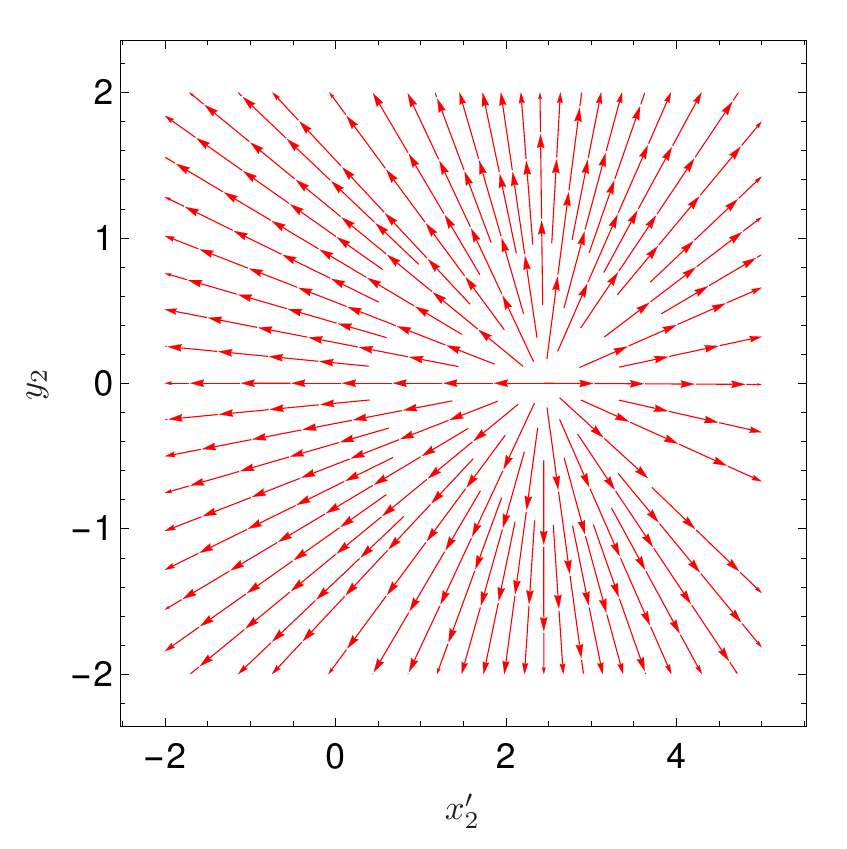}
	\caption{ Streamplots of the probability currents $ {\bf j}_{1} \bigl((x'_1,y_1), {\bf z_{2}}=0\bigr)$ (left panel) - a vector with components $j_{1,x}\bigl((x'_1,y_1),(0,0)\bigr)$ and $j_{1,y}\bigl((x'_1,y_1),(0,0)\bigr)$, and  $ {\bf j}_{2} \bigl({\bf z_{1}=0,(x_2',y_2)\bigr)}$ (right panel) - a vector with components $j_{2,x}\bigl((0,0),(x'_2,y_2)\bigr)$ and $j_{2,y}\bigl((0,0),(x'_2,y_2)\bigr)$. Here, $x'_1=x_1- x_0$ and $x'_2=x_2+x_0$ are the $x$-coordinates in the laboratory reference frame (see Fig.~\ref{fig:sketch}). The values of the parameters are : $\gamma=1$, $u=1/2$, $x_0=1$, $T_1=1$ and $T_2=2$. }
	\label{fig:curr}
\end{figure}

%We have shown in the previous section that no permanent rotational motion  of individual particle exist in the stationary %state, but a synchronization which leads to a non zero value of the absolute value of the curls. This seems to be %paradoxical because the one-dimensional model \cite{Li2019} was a Brownian gyrator \alb{Not sure I understand why this %is a paradox}. However, our model has  four degrees of freedom and the examination of the stream plots of the currents %highlight the complexity of the system.

Figure ~\eqref{fig:curr} shows streamplots of the currents  ${\bf j}_{1} \bigl((x_1,y_1), {\bf z_{2}}=0\bigl)$, i.e., the current associated with the particle $1$ with the particle $2$ being fixed at the center of its optical trap, and  $ {\bf j}_{2} \bigl({\bf z_{1}}=0,(x_2,y_2))$ - the current associated with particle $2$ with the particle $1$ being fixed at the center of its trap. 
 In this and the subsequent figure we choose the following values of the parameters :  $\gamma=1$,  $u=1/2$, $x_0=1$, $T_1=1$ and $T_2=2$. 
We observe that the streamplot of ${\bf j}_{1} \bigl((x_1,y_1), {\bf z_{2}}=0)$ consists of curves which travel from infinity to some fixed point, while the one for $ {\bf j}_{2} \bigl({\bf z_{1}}=0,(x_2,y_2))$ consists of curves which starts from some point and travel to infinity.  In both cases the curves are not closed, as it happens for the Brownian gyrator (see e.g. \cite{Cerasoli2018}); the reason for such a behavior is that both components of each current are living at the same temperature.

\begin{figure}[h]
	\center
	%\psfrag{ }[ct][ct][1.]{ }
	\includegraphics[width=5.8cm]{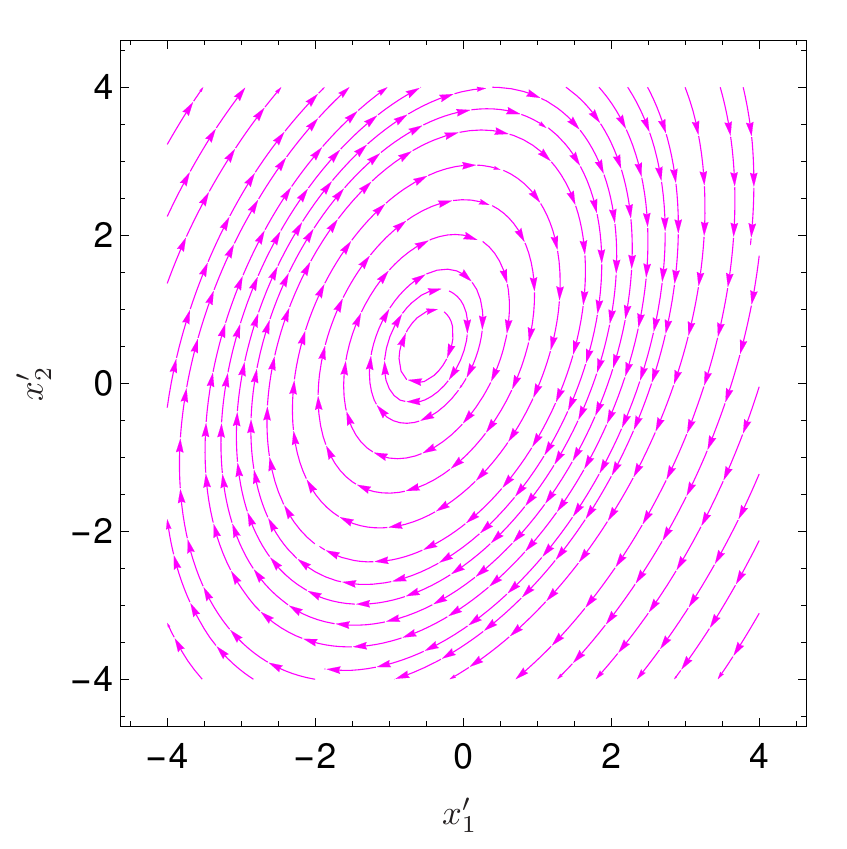}
	\hspace{.75cm}\includegraphics[width=5.8cm]{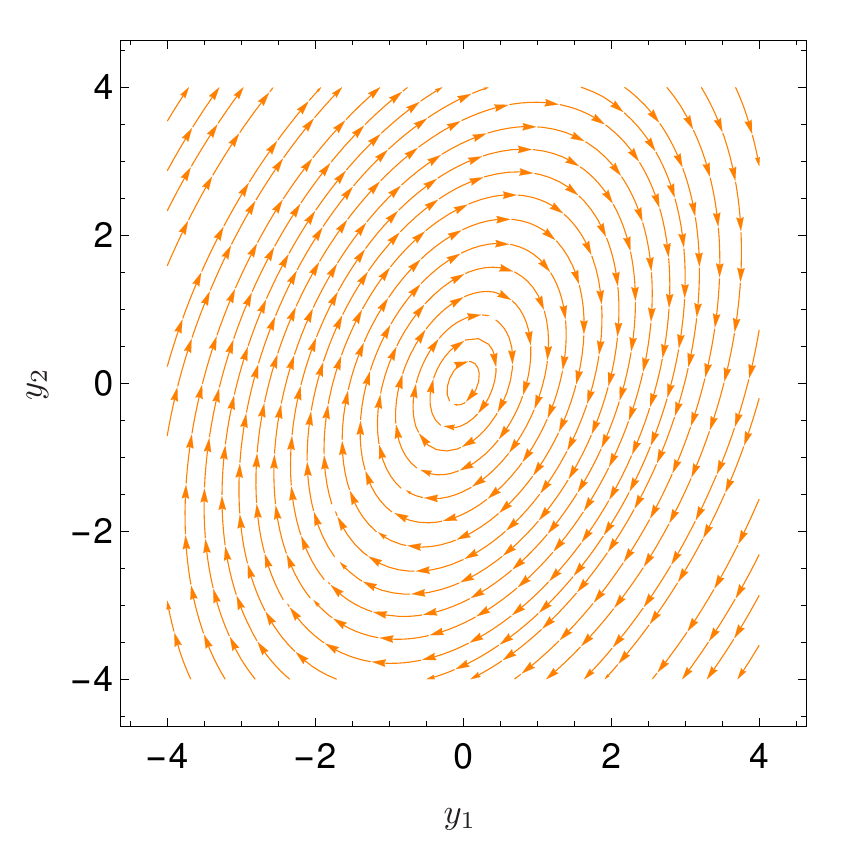}
	\caption{Streamplots of the  vectors $ \bigl(j_{1,x} \bigl({\bf z_{1}}, {\bf z_{2}}\bigr),j_{2,x} \bigl({\bf z_{1}}, {\bf z_{2}}\bigr)\bigr) $ with fixed $y_1=1, \, y_2=2$ (left panel),  and   $ \bigl(j_{1,y} \bigl({\bf z_{1}}, {\bf z_{2}}\bigr),j_{2,y} \bigl({\bf z_{1}}, {\bf z_{2}}\bigr)\bigr) $  with $x'_1=1, \, x'_2=2$ (right panel),  where $x'_1=x_1-x_0$ and $x'_2=x_2+x_0$ are the $x$-coordinates in the absolute reference frame (see Fig.~\ref{fig:sketch}). Here, $\gamma=1$, $u=1/2$, $x_0=1$, $T_1=1$ and $T_2=2$. }
	\label{fig:curr2}
\end{figure}

We consider next the behavior
of the $x$-components of the two currents,  which are subject to two \textit{different} temperatures, as well as the behavior of the $y$-components.
In the left panel in Fig.~\eqref{fig:curr2} we present a streamplot of  the vector with components $ \bigl(j_{1,x} \bigl({\bf z_{1}}, {\bf z_{2}}\bigr),j_{2,x} \bigl({\bf z_{1}}, {\bf z_{2}}\bigr)\bigr)$. Observe that the behavior is completely different from the one presented in Fig. \ref{fig:curr} -  the $x$-components of the two currents perform a circulation on the $(x_1,x_2)$-plane
along closed elliptic curves. Essentially the same behavior, which reveals an emerging cooperativity, is exhibited by the $y$-components of the two currents as depicted on the left panel in Fig.~\eqref{fig:curr2}. This is precisely what was previously observed for the Brownian gyrator model on a plane with different temperatures along the two Cartesian directions. Here, however, neither of the particles themselves performs a gyration along some point on a plane but rather 
the components of the currents of two particles circulate along closed orbits in a correlated manner.

%
%\begin{figure}[h]
%	\center
%	%\psfrag{ }[ct][ct][1.]{ }
%	\includegraphics[width=6cm]{plotj1xj2x.pdf}
%	\includegraphics[width=6cm]{plotj1yj2y.pdf}
%	\caption{ Stream plots of the probability currents $ (j_{1,x} \bigl({\bf z_{1}}, {\bf z_{2}}=0),j_{2,x} \bigl({\bf z_{1}}, {\bf z_{2}}=0)) $ (left panel) and   $ (j_{1,y} \bigl({\bf z_{1}}, {\bf z_{2}}=0),j_{2,y} \bigl({\bf z_{1}}, {\bf z_{2}}=0)) $ (right panel),  with ${\bf z_{1}}=(x_1,y_1)$ and  with $\gamma=1$, $u=1/2$, $r=1$, $T_1=1$, $T_2=2$. }
%	\label{fig:curr1}
%\end{figure}

To further characterise the circulation of the probability currents on the planes $(x_1,x_2)$ and $(y_1,y_2)$, evidenced in Fig. \ref{fig:curr2}, we evaluate below several additional properties. These are a) the values of the curls of the probability currents 
at positions of the optical traps, 
b) the mean angular momenta and c) the mean angular velocities of the circulation. 

a) Consider first the curls of the probability currents on the planes $(x_1,x_2)$ and $(y_1,y_2)$ defined as
\begin{align}
	S_{x}\bigl({\bf z_{1}},{\bf z_{2}}\bigr) =&
  \; 
\frac{\partial}{\partial x_{1}} j_{2,x}\bigl({\bf z_{1}},{\bf z_{2}}\bigr) \; - \; 
\frac{\partial}{\partial x_{2}} j_{1,x}\bigl({\bf z_{1}},{\bf z_{2}}\bigr)
\end{align}
and
\begin{align}
	S_{y}\bigl({\bf z_{1}},{\bf z_{2}}\bigr) =&
	\; 
	\frac{\partial}{\partial y_{1}} j_{2,y}\bigl({\bf z_{1}},{\bf z_{2}}\bigr) \; - \; 
	\frac{\partial}{\partial y_{2}} j_{1,y}\bigl({\bf z_{1}},{\bf z_{2}}\bigr) \,.
\end{align}
%We set next $x_1=x_2=0$ (i.e. the $x$-components are fixed to the centers of the respective optical traps) and define the %properties  integrated over the $y$-components
%\begin{equation}
%\overline{S_x}=\int dy_1 \int dy_2 \, S_x\bigl(\{0,y_1\},\{0,y_2\}\bigl) \,.
%\end{equation}
%Similarly, to characterise the gyration on the plane $(y_1,y_2)$, we set $y_1=y_2=0$ and integrate over the $x$-components, which leads to
%\begin{equation}
%	\overline{S_y}=\int dx_1 \int dx_2 \, S_y\bigl(\{x_1,0\},\{x_2,0\}\bigr) \,.
%\end{equation}
Taking advantage of Eqs. \eqref{eq:j1}, we find that the values of these curls at the locations of the centers of the optical traps are given explicitly by
\begin{equation}
\begin{split}
\label{Si}
	S_x(\{0,y_1\},\{0,y_2\})&= \frac{2u\kappa\Delta (T_1-T_2)}{((T_1 - T_2)^2 u^2 + 4 T_1 T_2\kappa^2)^2} P\bigl(\{0,y_1\}, \{0,y_2\}\bigr) \\
	S_y(\{x_1,0\},\{x_2,0\})&=\frac{2u\kappa^2(T_2^2-T_1^2)}{(T_1 - T_2)^2 u^2 + 4 T_1 T_2\kappa^2}
	 P\bigl(\{x_1,0\}, \{x_2,0,\}\bigr)
	 \end{split}
\end{equation}
where 
\begin{equation}\label{eq:delta}
\begin{split}
	\Delta &=8 x_0^2 (T_1 - T_2)^2 u^4 - (T_1 - T_2)^2 u^2 (T_1 + T_2 + 
	16 x_0^2 u)\kappa \\&+ 16 x_0^2 (T_1^2 + T_2^2) u^2\kappa^2 - 
	4 T_1 T_2 (T_1 + T_2)\kappa^3 \,.
	\end{split}
\end{equation}
Note that there is no symmetry between the expressions in the first and the second line in Eqs. \eqref{Si}, which is due to the fact 
that the centers of both optical traps are located on the $x$-axis. 

%It is worth noting that Eq.\ref{eq:delta} is finite when $u$ goes to zero \alb{Why is this relevant here? Haven't we already written that the curls vanish for $u=0$?}. Moreover, $S_x(\{0,y_1\},\{0,y_2\})$ is simply proportional to 
% $P\bigl(\{0,y_1\}, \{0,y_2\}\bigr)$ and averaging over the $y$-components leads to the  proportionality factor $u(T_1-%T_2)$. Similarly, in the subspace $(y_1,y_2)$,  $S_x(\{0,y_1\},\{0,y_2\})$ is simply proportional to 
% $P\bigl(\{0,y_1\}, \{0,y_2\}\bigr)$ . 
 
% The existence of a non zero values of the curls of the current can be also characterized by the angular momentum of the two subspaces. 
b) The angular momentum (per unit mass) for the rotation of the probability current 
 on the $(x_1,x_2)$-plane is defined as

\begin{equation}
L_{x_1,x_2}=  x_1 j_{2,x} -x_2 j_{1,x} \,, 
\end{equation}
such that its averaged value is given by 
\begin{equation}
\average{L_ {x_1,x_2}}=\int \int d{\bf z}_1  d{\bf z_2}  \, \left( x_1 j_{2,x} -x_2 j_{1,x}\right) \,.
\end{equation}

Similarly,  
 the angular momentum for the rotation on the $(y_1,y_2)$-plane and its averaged value 
 follow 

 \begin{equation}
	L_{y_1,y_2}= y_1 j_{2,y} - y_2 j_{1,y} \,,
\end{equation}
and
\begin{equation}
	\average{L_ {y_1,y_2}}=\int \int d{\bf z}_1 d{\bf z}_2 \, \left( y_1 j_{2,y} - y_2 j_{1,y}\right).
\end{equation}

\begin{figure}[t]
	\includegraphics[width=8cm]{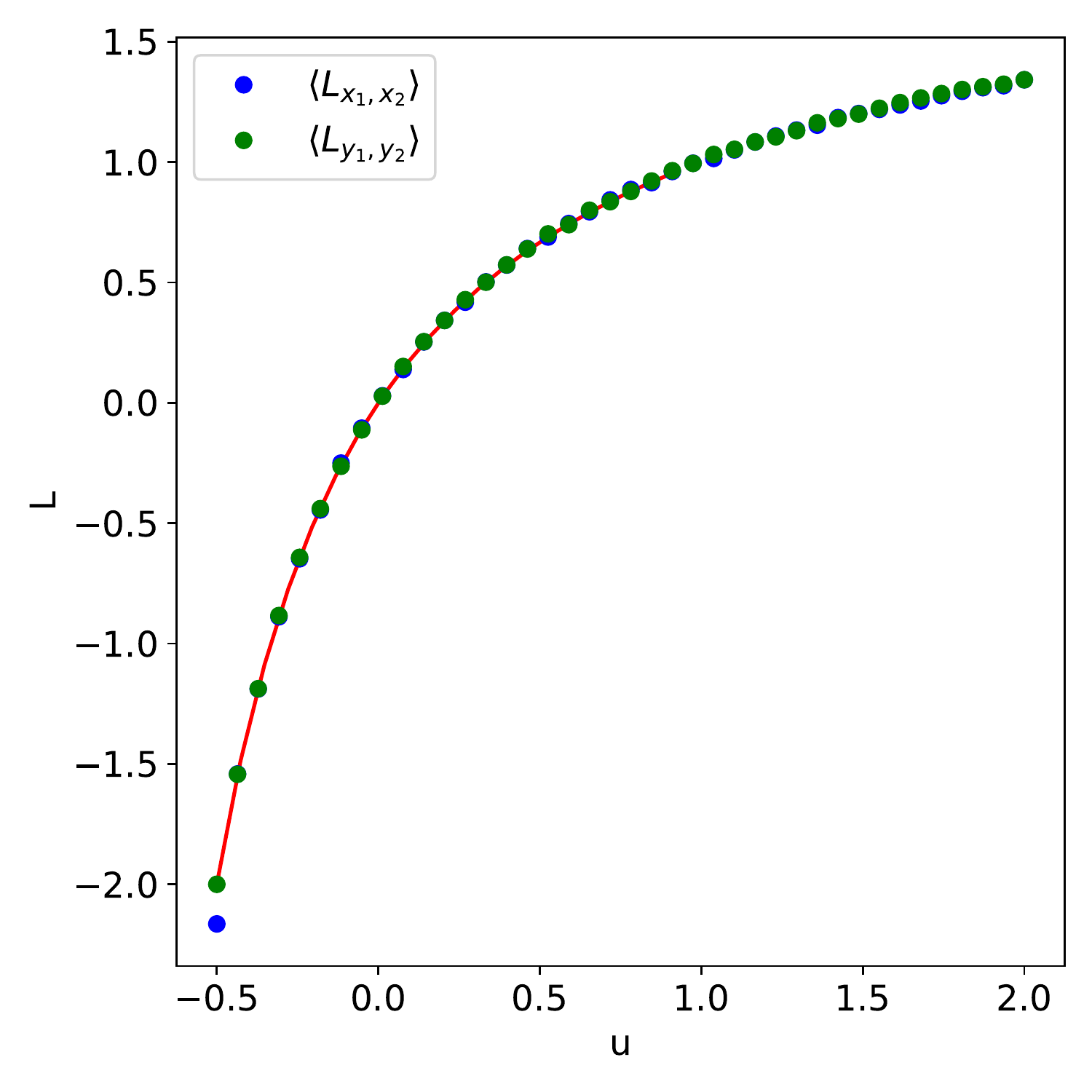}
         \includegraphics[width=8cm]{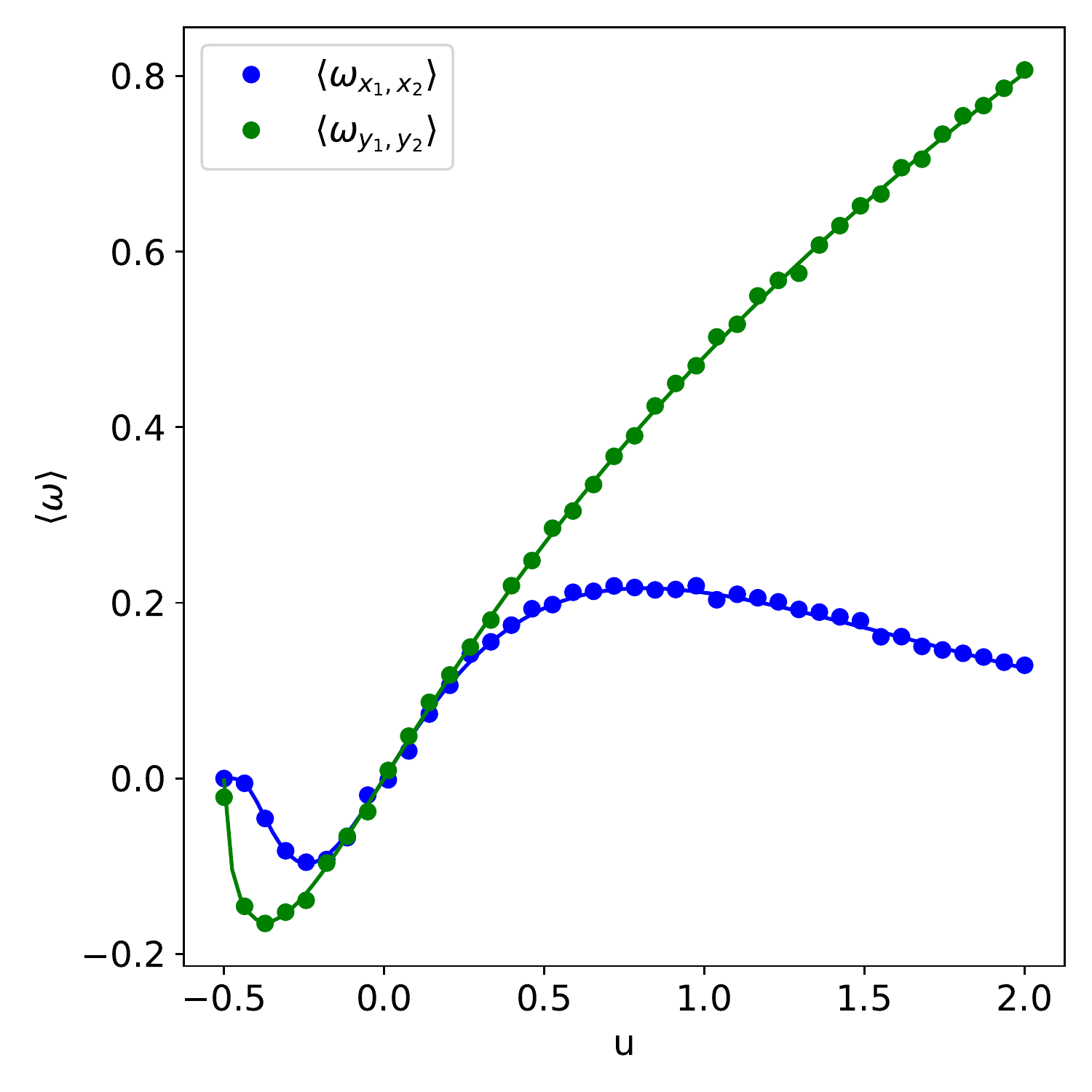}
	\caption{ Left panel: Avraged angular momenta  $\average{L_{x_1,x_2}}$ and $\average{L_{y_1,y_2}}$ as functions of  the coupling parameter $u$. Numerical results are given by the full circles, while the analytical expression \eqref{eq:Lxx} is given by the solid curve.
 Right panel: Averaged angular velocities  $\average{\omega_{x_1,x_2}}$ and $\average{\omega_{y_1,y_2}}$  as functions of  $u$. Numerical results are denoted by full circles and the analytical expressions \eqref{eq:Omegaxx} are denoted by solid curves. In both panels $\gamma=1$, $T_1=1$ and $T_2=3$.}
	\label{fig:model}
\end{figure}

Using our Eqs.~\eqref{eq:j1} and performing the integrals, we find that the averaged
values of the angular momenta on the $(x_1,x_2)$ and $(y_1,y_2)$ are exactly equal to each other 
and are both given by a very simple expression
\begin{equation}\label{eq:Lxx}
	\average{L_{x_1,x_2}}=	\average{L_{y_1,y_2}}=\frac{u(T_2-T_1)}{\kappa}.
\end{equation}
The equality of both averaged angular momenta is rather surprising in view of the fact that the values of the curls on these planes are very different - see Eqs. \eqref{Si}. The averaged angular momenta defined in Eq. \eqref{eq:Lxx}  
  are depicted on the left panel in Fig. \ref{fig:model} as functions of the coupling parameter 
$u$, (recall that $\kappa = u + \gamma$). The prediction in Eq. \eqref{eq:Lxx} is confirmed by 
 numerical simulations  by using an Euler-Maruyama method \cite{Kloeden_Platen} with  
 time step
$\delta t=2\times 10^{-3}$ and the total elapsed time 
$t_f=2000$. Each simulation result (given by filled circles in Fig. \ref{fig:model})  
is an average performed over $48$ to
$96$ independent runs. 

%As expected the angular momentum cancels when the coupling between the two particles vanishes and when the %temperatures are equal, which corresponds to a equilibrium state, in which no stationary current appears. When the %interaction between particles is repulsive and stable, ($-\kappa<u<0$),  the mean angular velocity is negative and the %lower bound is given by $T_1-T_2$. When the interaction is attractive, the angular velocity is positive and when the %coupling strength $u$ becomes large, the angular velocity goes to an asymptotic value $T_2-T_1$, which is the opposite %of the value obtained for repulsive interaction.  

c) Lastly, 
we calculate the averaged angular velocities $\average{\omega_ {x_1,x_2}}$ and  $\average{\omega_ {y_1,y_2}}$ of circulations of the probability currents on the planes $(x_1,x_2)$ and $(y_1,y_2)$ which are defined
as 
%\begin{equation}
%	\omega_{x_1,x_2}=\frac{x_1v^x_2-x_2v^x_1}{x_1^2+x^2_2},
%\end{equation}
%in the subspace $(x_1,x_2)$ and as 
%\begin{equation}
%	\omega_{y_1,y_2}=\frac{y_1v^y_2-y_2v^y_1}{y_1^2+y^2_2},
%\end{equation}
%in the subspace $(y_1,y_2)$.
%The average of the mean velocities $\average{\omega_{x_1,x_2}}$ and $\average{	\omega_{y_1,y_2}}$ are 
%expressed as
\begin{equation}
\begin{split}
	\average{\omega_ {x_1,x_2}}&=\int \int d{\bf z}_1 d{\bf z}_2 \, \left( \frac{x_1j_{2,x} -x_2 j_{1,x}}{x_1^2+x^2_2} \right),\\
	\average{\omega_ {y_1,y_2}}&=\int \int d{\bf z}_1  d{\bf z}_2 \, \left(\frac{y_1j^y_2-y_2j^y_1}{y_1^2+y^2_2} \right) .
	\end{split}
\end{equation}
Performing the integrals, we eventually find
\begin{equation}\label{eq:Omegaxx}
\begin{split}
	\average{\omega_{x_1,x_2}}& =	u(T_2-T_1)\sqrt{\frac{\kappa^2-u^2}{(T_1 - T_2)^2 u^2 + 4 T_1 T_2\kappa^2}}\\&\times \exp\left(-\frac{8x_0^2(T_1+T_2)u^2\kappa^2}{(u+\kappa)((T_2-T_1)^2u^2+4T_1T_2\kappa^2)}\right), \\
	\average{\omega_{y_1,y_2}}&=	u(T_2-T_1)\sqrt{\frac{\kappa^2-u^2}{(T_1 - T_2)^2 u^2 + 4 T_1 T_2\kappa^2}}.
	\end{split}
\end{equation}
Equation \eqref{eq:Omegaxx} implies that  the ratio of the averaged angular velocities obeys
\begin{equation}
\frac{\average{\omega_{x_1,x_2}}}{\average{\omega_{y_1,y_2}}} = \exp\left(-\frac{8x_0^2(T_1+T_2)u^2\kappa^2}{(u+\kappa)((T_2-T_1)^2u^2+4T_1T_2\kappa^2)}\right)  < 1\,,
\end{equation}
i.e., this ratio is always less than unity, despite the fact that the averaged angular momenta and correspondingly, the torques are equal to each other (see Eq. \eqref{eq:Lxx}). Consequently, the averaged angular velocity for the rotations on the $(y_1,y_2)$-plane is always greater than the one for the rotations on the $(x_1,x_2)$-plane, for an arbitrary sign of the coupling parameter $u$. 

 Overall, $\average{\omega_{y_1,y_2}}$ and $\average{\omega_{x_1,x_2}}$ 
 are non-monotonic functions of the parameter $u$ with a minimum attained at some $u = u^* < 0$. For large positive values of $u$ the behavior of   $\average{\omega_{y_1,y_2}}$ and $\average{\omega_{x_1,x_2}}$ is markedly different:
  $\average{\omega_{y_1,y_2}}$ diverges in proportion to a  square-root of $u$:
\begin{equation}\label{eq:Omegayyasym}
	\average{\omega_{y_1,y_2}}\simeq	\frac{T_2-T_1}{T_2+T_1} \sqrt{\gamma \, u} \,,
\end{equation}
 while $\average{\omega_{x_1,x_2}}$ attains a maximal value when you $u$ approaches
 \begin{equation}
 u = \frac{(T_1 - T_2)^2}{8 x_0^2 (T_1+T_2)}
 \end{equation}
 and then \textit{decreases} exponentially, 
\begin{equation}
\label{eq:Omegaxxasym}
	\average{\omega_{x_1,x_2}}\simeq	\frac{T_2-T_1}{T_2+T_1} \sqrt{\gamma \, u}
	\exp\left(-\frac{ 4 x_0^2 (T_1+T_2) \, u}{(T_2 - T_1)^2}\right) \,.
\end{equation}
The behavior of  $\average{\omega_{y_1,y_2}}$ and $\average{\omega_{x_1,x_2}}$ as functions of $u$ is depicted on the right panel in Fig. \ref{fig:model} together with the results of numerical simulations which confirm our analytical predictions.

%which goes to zero when $u$ goes to $\infty$.
%Conversely, 
%\begin{equation}\label{eq:Omegayyasym}
%	\average{\omega_{y_1,y_2}}\simeq	\sqrt{u\gamma}\frac{T_2-T_1}{T_2+T_1}
%\end{equation}
%and diverges as the $\sqrt{u}$.

%The left panel of Fig.~\ref{fig:model} displays the mean value of the angular momentum $\average{L_{x_1,x_2}}$ and 
 %$\average{L_{y_1,y_2}}  $ versus $u$, as given by Eq.\eqref{eq:Lxx} together with the results of the numerical simulations. One obtains a perfect agreement with the exact result, Eq.\eqref{eq:Lxx}.
%The right panel of Fig.~\ref{fig:model}  shows the mean value of the angular velocities $\omega_{x_1,x_2}$ and $\omega_{y_1,y_2}$ as functions of $u$. Also in this case one obtains a perfect agreement between the numerical and the exact results. Whereas the curves overlap in the limit of small coupling $u$, one observes strong deviations for positive and negative values of $u$. This illustrates the fact that the rotation cannot be viewed as a solid body rotation for which angular momentum and angular velocity are proportional.
%\alb{Isn't this a consequence of the x-y asymmetry, i.e. that the traps are on the $x$-axis?}

\section{Conclusion}\label{sec:conclusion}

To conclude, we presented here a 
detailed theoretical analysis of an out-of-equilibrium dynamics of two interacting, randomly moving particles in a two-dimensional system, 
which was realised experimentally in \cite{Berut2014,Berut2016}. More specifically, the experimental set-up in these references consisted of a disc-shaped shallow cell filled with a solvent and containing 
two suspended micrometer-sized beads, each being held by its own optical tweezer.  One of the tweezers was subject to an additional, externally-imposed noise such that the particle held by this very tweezer lived 
at an effectively different temperature as compared to the other one.  
Due to the presence of a solvent, the particles were coupled by hydrodynamic interactions.  

References \cite{Berut2014,Berut2016} focused on the behavior of the effective heat fluxes between the two beads in the out-of-equilibrium state with unequal temperatures and developed both experimental and theoretical analyses.  On the theoretical side, the dynamics was framed
in terms 
 of two coupled over-damped Langevin equations with effective harmonic interactions between the particles, 
 the parameters of which were deduced from 
 the Rotne-Prager diffusion tensor.  It was demonstrated that the heat fluxes obey, e.g.,
  an exchange fluctuation theorem which result was confirmed both experimentally and theoretically, with 
  a very good agreement  between the two approaches. In turn, it proved directly the validity of the theoretical description based on the Langevin dynamics.
  
On the other hand, the theoretical analysis in  \cite{Berut2014,Berut2016} was based on the assumption that the stochastic dynamics of the two particles can be viewed as an effectively \textit{one-dimensional} process that evolves along the line connecting the centers of two tweezers. Here, we addressed a conceptually important question what physical effects can be potentially overlooked due to such an assumption. To this end, we formulated and analysed essentially the same model but with two particles evolving on a \textit{plane}, which is in fact closer to the actual experimental  set-up.

We have shown that, indeed, a reduction of the dynamics to a one-dimension misses some rather spectacular effects. 
We demonstrated that in case when the temperatures at which the particles live are different, the system reaches a steady-state with non-zero probability currents which possess non-zero curls.
 As a consequence, in such a system the particles are continuously spinning around their centers of mass in a completely synchronised way 
 - the curls of currents at the instantaneous positions of two particles have the same magnitude and sign.
 Further on, our analysis revealed emerging correlations between the probability currents. In particular, we realised that 
 the $x$- components (and also the $y$-components) of the currents  undergo  a rotational motion along closed elliptic orbits. 
 
 \section*{Acknowledgments} 
 
 The authors wish to thank Luca Peliti for many helpful discussions.

\begin{appendix}
\section{Small coupling limit}
\label{ap}

In this  appendix we  focus on  the  behavior in the limit of a  vanishingly small coupling parameter $u$, in which case our results attain very simple forms. For $u \to 0$, 
Eqs. \eqref{11} and \eqref{12} become
\begin{eqnarray}
	\label{24}
	{\bf j}_{1} \bigl({\bf z_{1}}, {\bf z_{2}}\bigr)  &\simeq&
	u \, \frac{T_{1} - T_{2}}{2 T_{2}} \,  {\bf z_{2}} 
	\;  P\bigl({\bf z_{1}}, {\bf z_{2}}\bigr) \,,
	\\
	\nonumber
	\\
	\label{25}
	{\bf j}_{2} \bigl({\bf z_{1}}, {\bf z_{2}}\bigr)  &\simeq&
	u \, \frac{T_{2} - T_{1}}{2 T_{1}} \,  {\bf z_{1}} 
	\;  P\bigl({\bf z_{1}}, {\bf z_{2}}\bigr) \,,
\end{eqnarray}
and the probability density function $P\bigl({\bf z_{1}}, {\bf z_{2}}\bigr)$ in Eq. \eqref{9} attains the form
\begin{equation}
	\label{26}
	P\bigl({\bf z_{1}}, {\bf z_{2}}\bigr) \; \simeq \; 
	Z^{-1} \, 
	\exp\biggl(
	-\frac{\gamma}{2 T_{1}} \, z_{1}^{2} \; - \; \frac{\gamma}{2 T_{2}} \, z_{2}^{2} 
	\; + \; u \frac{(T_{1}+T_{2})}{2T_{1}T_{2}} \, ({\bf z_{1}} \cdot {\bf z_{2}}) 
	\; + \; \frac{2u}{T_{1}} \, ({\bf r}  \cdot {\bf z}_{1}) \; - \; \frac{2u}{T_{2}} \, ({\bf r} \cdot {\bf z}_{2})
	\biggr)
\end{equation}
with
\begin{equation}
	\label{27}
	Z \; \simeq \; \frac{4 \pi^{2} T_{1} T_{2}}{\gamma^{2}} \,.
\end{equation}
Using Eqs. \eqref{24} and \eqref{25}, 
one readily calculates the curls of the probability currents  to get
\begin{eqnarray}
	\nonumber
	S_{1}\bigl({\bf z_{1}},{\bf z_{2}}\bigr) &=&
	{\boldsymbol \nabla}_{1}\times {\bf j}_{1} \; = \; 
	\frac{\partial}{\partial x_{1}} j_{1,y} \; - \; 
	\frac{\partial}{\partial y_{1}} j_{1,x}
	\\
	\nonumber
	\\
	\nonumber
	&=&
	u \, \frac{T_{1} - T_{2}}{2 T_{2}} \,
	\Bigl(
	y_{2} \frac{\partial}{\partial x_{1}} P\bigl({\bf z_{1}}, {\bf z_{2}}\bigr) 
	\; - \; 
	x_{2} \frac{\partial}{\partial y_{1}} P\bigl({\bf z_{1}}, {\bf z_{2}}\bigr) 
	\Bigr)
	\\
	\nonumber
	\\
	&\simeq&
	u \, \frac{T_{1} - T_{2}}{2 T_{1} T_{2}} \,
	\Bigl(
	2u \, \bigl({\bf r}\times {\bf z}_{2}\bigr) \; - \; \gamma \, \bigl({\bf z}_{1}\times {\bf z}_{2}\bigr)
	\Bigr)
	P\bigl({\bf z_{1}}, {\bf z_{2}}\bigr) 
	\label{28}
\end{eqnarray}
The above expression simplifies considerably in case when the particle $1$ resides in the center
of its optical trap, i.e. 
 ${\bf z}_{1} = 0$, 
 \begin{equation}
	\label{29}
	S_{1}\bigl(0,{\bf z_{2}}\bigr) 
	\; \simeq \; 
	u^{2} \, \frac{(T_{1} - T_{2})}{T_{1} T_{2}} \, \bigl({\bf r}\times {\bf z}_{2}\bigr) \, 
	P\bigl(0, {\bf z_{2}}\bigr) \,,
\end{equation}
or explicitly,
\begin{equation}
	\label{30}
	S_{1}\bigl(0,{\bf z_{2}}\bigr) 
	\; \simeq \; 
	\frac{u^{2} \gamma^{2}}{4\pi^{2}}  \, \frac{(T_{1} - T_{2})}{T_{1}^{2} T_{2}^{2}} \, 
	\bigl({\bf r}\times {\bf z}_{2}\bigr) \, 
	\exp\biggl(
	-\frac{\gamma}{2 T_{2}} \, z_{2}^{2}
	\; - \;
	\frac{2u}{T_{2}} \, ({\bf r} \cdot {\bf z}_{2})
	\biggr) \,.
\end{equation}
Equation \eqref{30} shows in a transparent way that the curl of the probability vanishes when the temperatures $T_1$ and $T_2$ are equal to each other, and also when the coupling parameter $u$ or the stiffness $\gamma$ of the optical trap are equal to zero.
% One can easily calculate
%the average of the absolute value of this spinning which is equal to $0$.
%
\end{appendix}

%\bibliography{model21.bib}

\end{document}